\documentclass[pra,letterpaper,aps,10pt,superscriptaddress,twocolumn,floatfix]{revtex4-1}
\usepackage{amsmath,graphicx,amssymb,braket,xcolor,upgreek}
\usepackage[FIGTOPCAP]{subfigure}
\usepackage[colorlinks, linkcolor=blue, citecolor=blue, urlcolor=blue, breaklinks=true]{hyperref}
\usepackage{empheq,verbatimbox}

\begin{document}

\title{A realization of a quasi-random walk for atoms in time-dependent optical potentials}
\author{Torsten Hinkel, Helmut Ritsch and Claudiu Genes}
\affiliation{Institut f\"ur Theoretische Physik, Universit\"at
Innsbruck, Technikerstrasse 25, A-6020 Innsbruck, Austria}
\date{\today}
\date{\today }

\begin{abstract}
We consider the time dependent dynamics of an atom in a two-color
pumped cavity, longitudinally through a side mirror and
transversally via direct driving of the atomic dipole. The beating
of the two driving frequencies leads to a time dependent effective
optical potential that forces the atom into a non-trivial motion,
strongly resembling a discrete random walk behavior between lattice
sites. We provide both numerical and analytical analysis of such a
quasi-random walk behavior.

\end{abstract}

\maketitle

\affiliation{Institute for Theoretical Physics, University of
Innsbruck, Technikerstrasse 25, A-6020 Innsbruck, Austria}

\section{Introduction}

Over the past few decades, the optical control of motional degrees
of freedom has seen great progress both on the experimental and
theoretical fronts. As a particular example of such achievement, the
cavity QED setting provides a paradigm for the observation and
manipulation of motion of atoms, ions or atomic ensembles via
tailored cavity
modes~\cite{Mabuchi1999Full,Hood2000The,Maunz2004Cavity,
Leibrandt2009Cavity,Schleier2011Optomechanical}. In such a system,
the time delay between the action of the field onto the atomic
system's motion via the induced optical potential and the
back-action of the particle's position change onto the cavity field
leads to effects such as
cooling~\cite{Horak1997Cavity,Hechenblaikner1998Cooling,Domokos2003Mechanical}
or self-oscillations (for a recent review
see~\cite{Ritsch2013Cold}). Addressing of the particle's motion
works best for quantum emitters with sharp transitions such as ions
or atoms but manipulation via the effective static polarizability is
also possible in the case of molecules in standing wave~\cite{Lev2008Prospects} or ring
cavities~\cite{Schulze2010Optomechanical}, or macroscopic particles
such as levitated dielectric micron-sized
spheres~\cite{Isart2011Levitated,Kiesel2013Cavity,Asenbaum2013Cavity}.
Typically, the driving is done either by direct pumping into the
cavity mode via one of the side mirrors (longitudinal pumping), or
indirectly via light scattering off the atom into the field mode
(transverse pumping). For transverse pumping, an exotic phenomenon
dubbed as self-organization can occur in the many atom
case~\cite{Domokos2002Collective}. Combinations of the two
techniques have been theoretically investigated in the limit of
equal laser frequencies~\cite{Niedenzu2013Seeding}; in such a case,
in a properly chosen rotating frame the effective combined optical
potential can be rendered time independent and the analysis greatly
simplifies.

\begin{figure}
  \includegraphics[width=0.9\columnwidth]{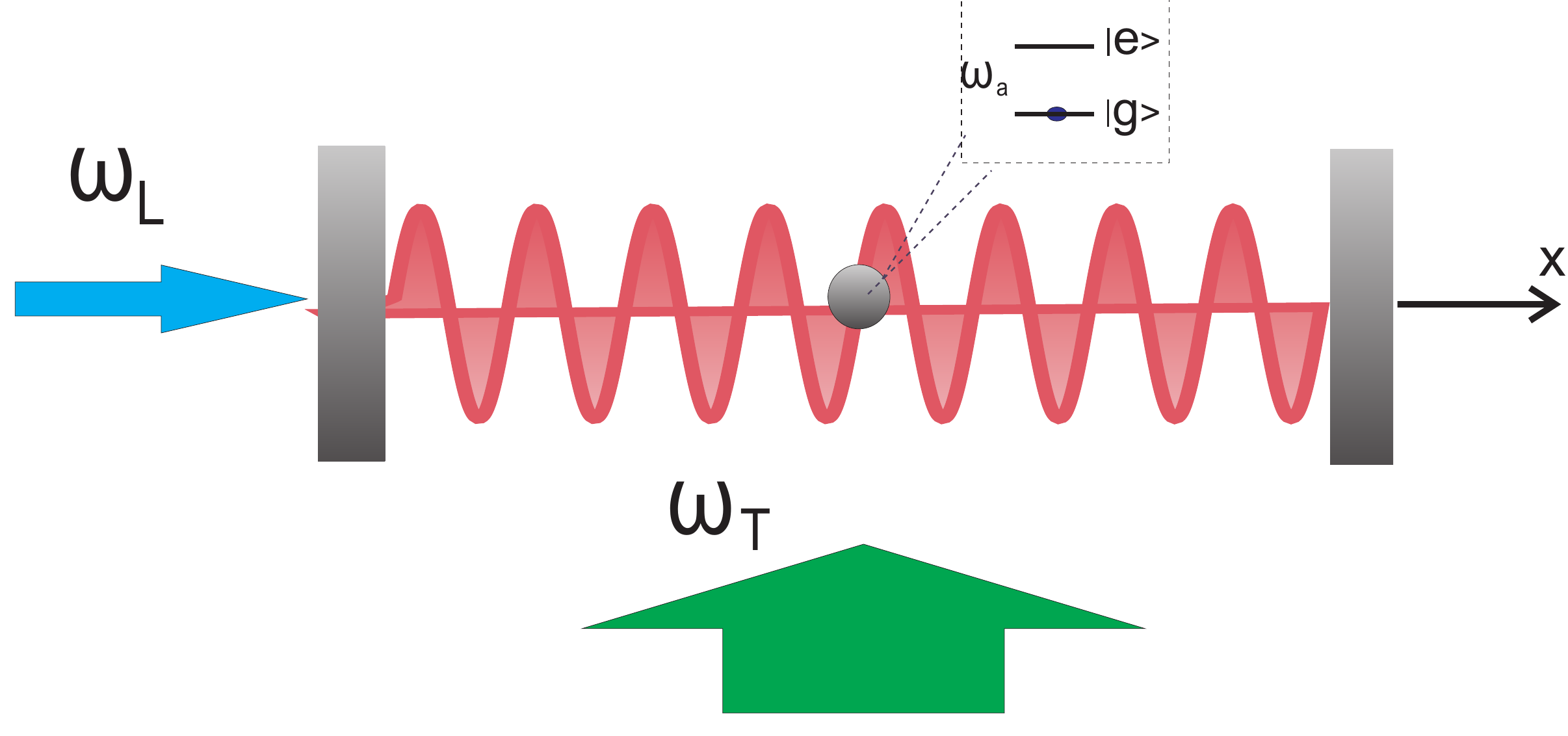}
  \caption{\textit{Schematics}. Illustration of the model considered involving an atom moving inside an optical
  cavity with longitudinal driving of strength $\eta_L$ and frequency $\omega_L$ and
  transverse driving $\eta_T$ and frequency $\omega_T$. The interference between the longitudinally pumped field
  and the field scattered from the transverse pump into the cavity mode leads to an effective optical potential that has a time dependent part oscillating at the
   frequency difference $\delta_{T}=\omega_L-\omega_T$.}
  \label{fig1}
\end{figure}

\noindent Here we depart from this scenario to consider frequency
beatings between the two pumps. While dynamics in the regime, where
each pump acts alone, is well understood, extra forces arise from
the interference between photons of different frequencies: i)
scattered from the transverse light field and ii) entering the
cavity mode from the longitudinal pump. The immediate effect of this
interference force is to generate a time-dependent optical potential
with a time modulation leading to a sign change that effectively
induces the particle into undergoing jumps along the cavity sites in
a quasi-random walk fashion. We analyze such a regime both
numerically and via simplified analytical models. The mechanism is
reminiscent of the one exploited in the creation of artificial
potentials in optical lattices~\cite{Struck2012Tunable} applied here
to the classical regime. By discretizing the trajectories, we
analyze the emerging discrete process via its correlation function
and find that it corresponds to an environment with a very short
memory. The goal of this analysis is to provide a quantum optical
setting in which the classical random walk can be observed and which
would constitute a starting point into further generalizations into
the quantum regime. In this sense, this work is a stepping stones
towards a proposal for implementing a quantum random walk mirroring
progress already achieved with
photons~\cite{Bouwmeester1999Optical}, atoms in optical
lattices~\cite{Duer2002Quantum}, ions in
traps~\cite{Travaglione2002Implementing} or on a one-dimensional
lattice of superconducting qubits~\cite{Ghosh2014Quantum}. Our
analysis is mainly based on a single two-level system but we discuss
as well an extension involving doped micro-spheres where the field
addresses a collective atomic variable (along the lines of hybrid
optomechanics with doped mechanical
resonators~\cite{Dantan2014Hybrid}).

The paper is organized as follows: in Sec.~II we introduce the
model. In Sec.~III we present numerical evidence showing the occurrence of a
quasi-random walk behavior and compute correlations of the engineered process that map close to those expected from a true random walk.
In Sec.~IV, we present a simplified analytical model that allows us
to derive the different forces acting on the particle and identify
different regimes and associated scalings for the occurrence of the
random walk. In Sec.~V, we discuss possible extensions of the model
involving a tailored driving via a frequency comb laser. We conclude
and present an outlook in Sec.~VI.

\section{Model}

We consider an effective one-dimensional model where an optical
cavity mode at $\omega _c$, decaying at rate $\kappa$ is driven
through a side mirror by a laser of amplitude $\eta _L$ and
frequency $\omega _L$. Transversally, a second laser drives the atom
directly with effective amplitude $\eta _{T}$ at $\omega _T$. The
longitudinal mode spatial variation inside the cavity is $f(x)=\cos
(k x)$ ($k$ is the corresponding wave-vector for the light mode with
wavelength $\lambda=2\pi/k$) and the atom-photon coupling is
specified by $g(x)=g f(x)$ (with $g$ being the maximum coupling).
The total system is described by the Hamiltonian
\begin{align}
    \hat{H} = \hat{H}_0+\hat{H}_p+\hat{H}_{JC},
\end{align}
consisting of a free part $\hat{H}_0$, a pumping term $\hat{H}_p$
and the Jaynes-Cummings interaction $\hat{H}_{JC}$. The free
evolution Hamiltonian describes the dynamics of a free particle of
mass $m$ and momentum operator $\hat{p}$ plus that of the cavity
mode (annihilation operator $\hat{a}$) and the two-level atom
\begin{equation}
    \hat{H}_0=\frac{\hat{p}^2}{2m}+\hbar \omega_{c} \hat{a}^\dagger \hat{a}+ \frac{\hbar \omega_{a}}{2} \hat{\sigma}^z.
\end{equation}
The atom dynamics is described with the help of the Pauli operators
$\hat{\sigma}^\pm$ and $\hat{\sigma}^z$ satisfying
$\left[\hat{\sigma}^+,\hat{\sigma}^-\right]=\hat{\sigma}^z$ and
$\left[\hat{\sigma}^\pm,\hat{\sigma}^z\right]=\mp2
\hat{\sigma}^\pm$. The atom's interaction with the cavity field is
included as a Jaynes-Cummings photon-excitation exchange process
quantified by the position dependent coupling strength $g f(x)$:
\begin{equation}
   \hat{H}_{JC}= i \hbar g f(x)(\hat{\sigma}^+\hat{a}-\hat{a}^\dagger \hat{\sigma}^-).
\end{equation}
Finally driving is included in the pump terms
\begin{equation}
    \begin{split}\hat{H}_{p}= & i \hbar \eta_L (\hat{a}^\dagger e^{-i\omega_L t}-\hat{a} e^{i\omega_L t})+\\
                        &i \hbar \eta_T (\hat{\sigma}^+ e^{-i\omega_T t}- \hat{\sigma}^- e^{i\omega_T
                        t}),
    \end{split}
\end{equation}
including direct pumping into mode $\hat{a}$ and atom driving of the
dipole operator $\hat{\sigma}^-$.

\begin{figure}[t]
  \includegraphics[width=0.8\columnwidth]{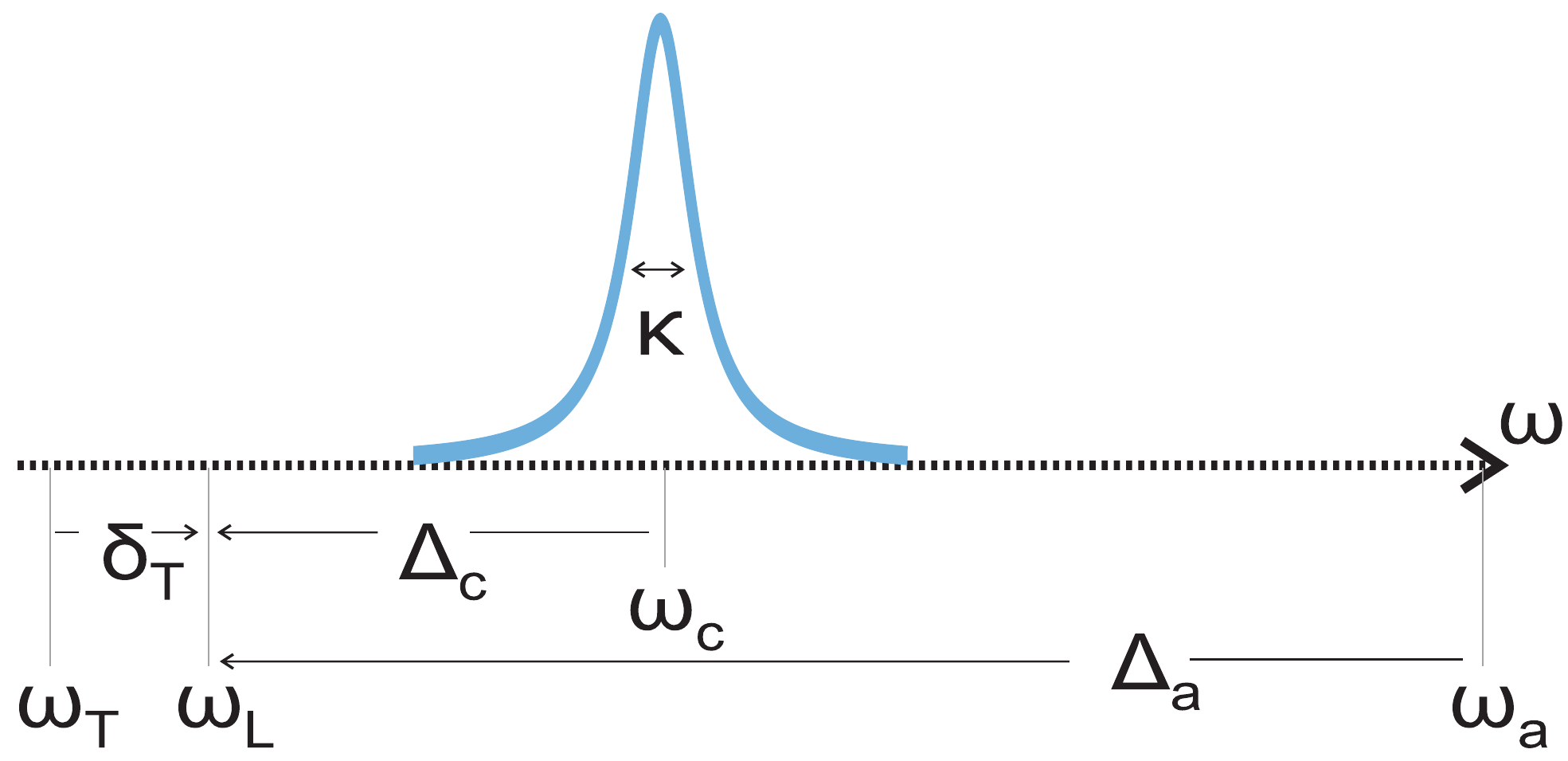}
  \caption{\textit{Frequencies}. Illustration of the frequencies and detunings as defined/used in the equations of motion. The detunings are taken with respect to the longitudinal driving frequency such that $\Delta_c<0$ corresponding to the stable regime of cavity QED with moving atoms (far from motional instability points) and $\Delta_a<0$ that corresponds to $U_0=g^2/\Delta_a<0$ as for high-field seekers.}
  \label{fig2}
\end{figure}

\noindent We proceed in a standard way to derive equations of motion
for classical quantities~\cite{Cohen1977quantum}. First, we make a
set of transformations to dimensionless normalized position $x=k
\langle \hat{x} \rangle$ and momentum $p=\langle
\hat{p}\rangle(\hbar k)^{-1}$. We denote the field amplitude by
$\alpha=\langle \hat{a} \rangle$ and the averaged atomic
polarization by $\beta=\langle \sigma^- \rangle$ (in a frame
rotating at $\omega_L$). In a first stage we consider finite
saturations of the population difference operator $\hat{\sigma}^z$
whose classical average we denote by $\beta^z$. We furthermore
assume that the build-up of quantum correlations between the atom
and the photon field can be neglected so that we can replace the
nonlinear terms such as $\hat{a} \hat{\sigma}^z $ by their
factorized classical averages $\alpha \beta^z$. The complete
equations of motion for atom and field are (including the
dissipative dynamics of the field mode at rate $\kappa$ and of the
atomic coherence at rate $\gamma$):
\begin{align}
\dot{\alpha } &=(-\kappa +i\Delta_c)\alpha
                -g f(x)\beta +\eta _L, \\
\dot{\beta }  &=(-\gamma +i\Delta_a)\beta
                -g f(x)\alpha \beta^z -\eta _T \beta^z e^{i\delta_T
                t},\\
\dot{\beta}^z  &=-2\gamma (\beta^z+1)
                -4 g f(x)\text{Re}\{\beta^* \alpha\} -4\eta _T \text{Re}\{ \beta^* e^{i\delta_T
                t}\}.
\end{align}
We introduced the detunings $\Delta_a=\omega_L-\omega_a$,
$\Delta_c=\omega_L-\omega_c$ and $\delta_T=\omega_L-\omega_T$
(illustrated in Fig.~\ref{fig2} with corresponding sign
conventions).

\begin{figure}[b]
  \includegraphics[width=0.99\columnwidth]{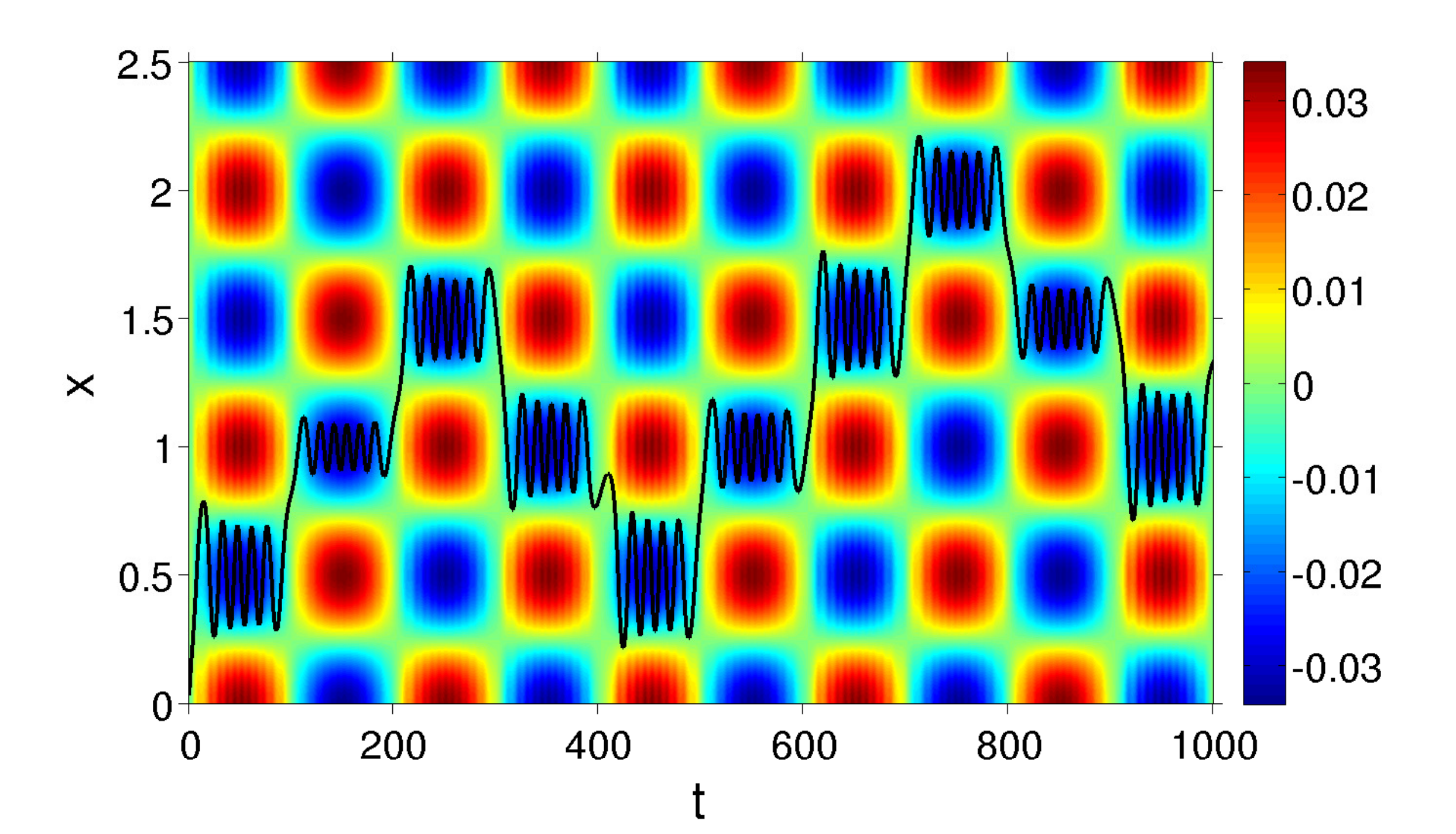}
  \caption{\textit{Quasi-random walk trajectory}. A single
  trajectory (black curve) overlapped with the effective potential
  as a function of increasing time for lattice sites appearing at
  full and half integer values of $x$ (with $\lambda=1$). The optical potential
  is illustrated as colored background and oscillates in time with a
  period of $T=\pi/\delta_{T}$. Note that $t$ is dimensionless as it is expressed in units of $\kappa^{-1}$ and $\kappa$ is set to unity.}
  \label{fig3}
\end{figure}

\noindent However, for the moment, we restrict our treatment to the
low saturation case, where $|\beta|^2\ll1$ which allows us to
linearize the $\hat{a}\sigma^z$ term by setting $\sigma^z\rightarrow
-1$. Such a linearized regime allows one to analytically derive the
forces acting on the particle. Numerical evidence points out that
this simplified limit provides similar effects with the finite
saturation case and we will base our analytical treatment on the
following simplified system of equations:
\begin{align}
\dot{\alpha } &=(-\kappa +i\Delta_c)\alpha
                -g f(x)\beta +\eta _L, \\
\dot{\beta }  &=(-\gamma +i\Delta_a)\beta
                +g f(x)\alpha +\eta _T e^{i\delta_T t}.
\label{eqalphabeta}
\end{align}
The motion of the atom is described by
\begin{align}
\dot{x} &= 2 \omega_{r} p, \\
\dot{p} &= -2 g f'(x) \text{Im}\{ \alpha^{\ast} \beta  \},
\label{eq:force1}
\end{align}
where we have condensed the particle's properties into the recoil
frequency $\omega_r=\hbar k^2/(2 m )$ and we have neglected
spontaneous emission induced momentum diffusion.

\begin{figure}[t]
 \subfigure[]{
  \includegraphics[width=0.99\columnwidth]{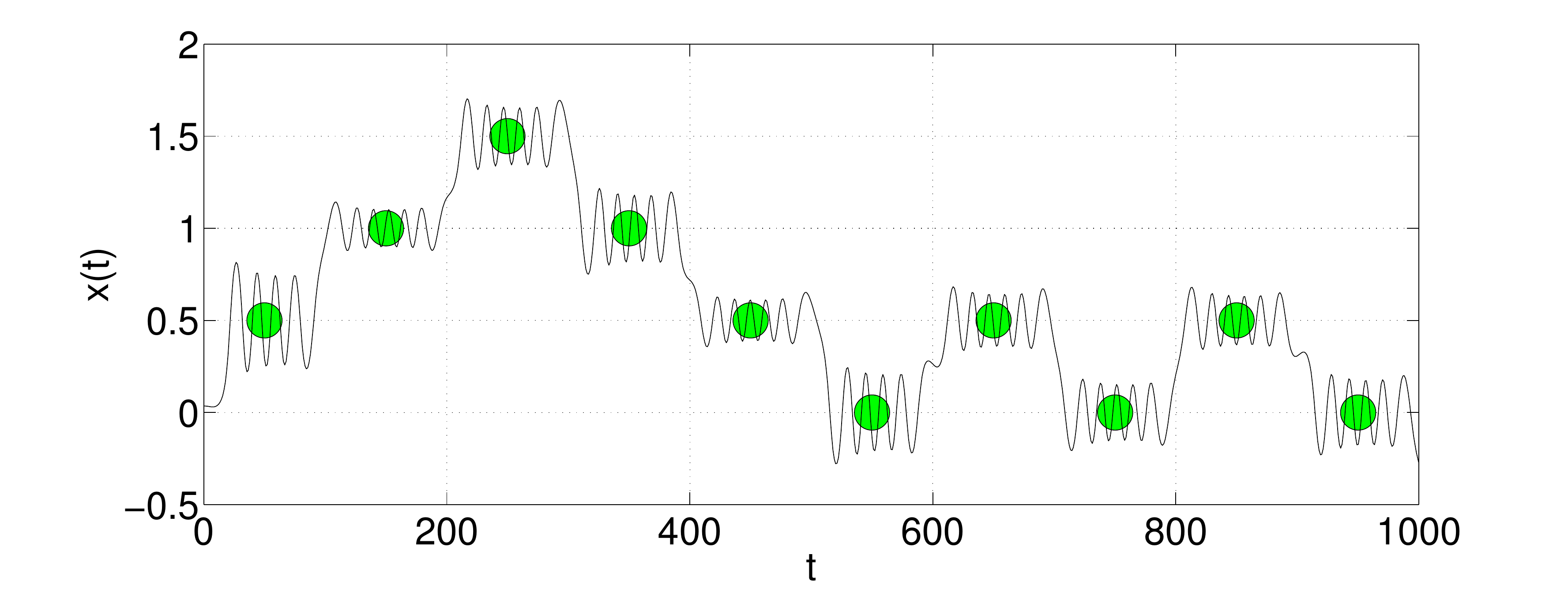}
  \label{fig4a}} \\
 \subfigure[]{
  \includegraphics[width=0.99\columnwidth]{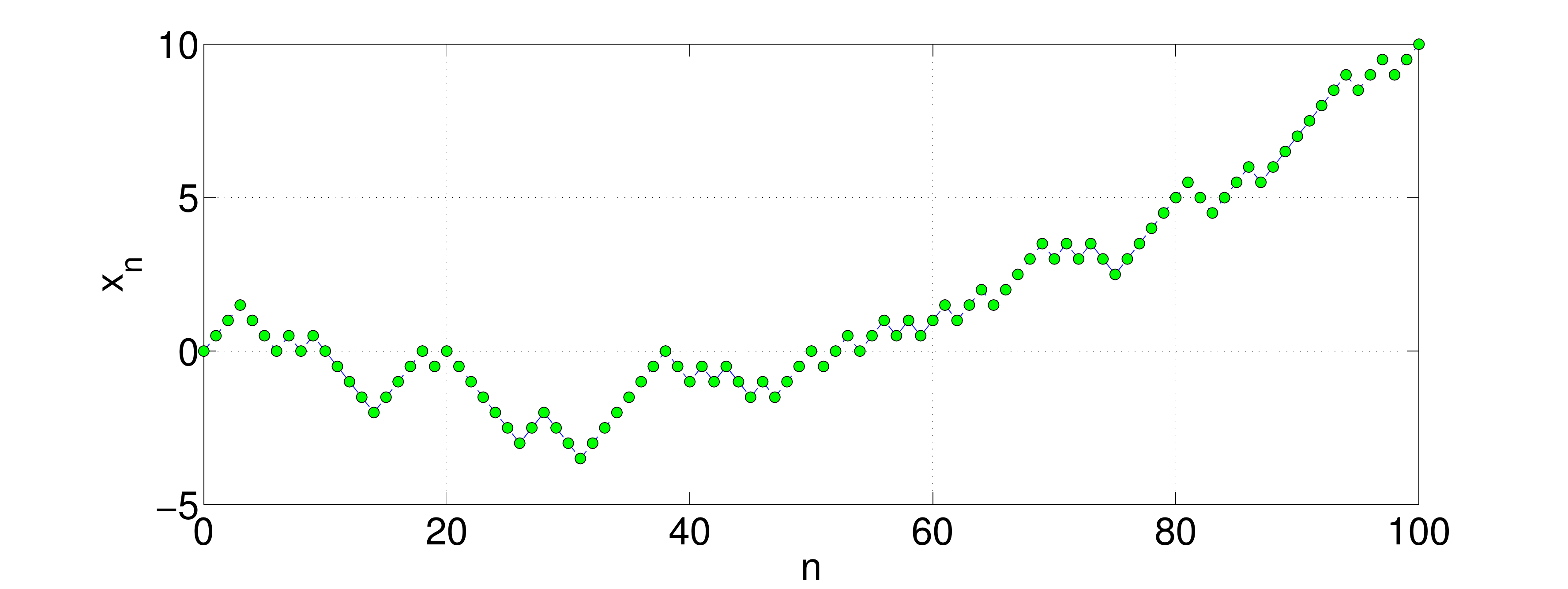}
  \label{fig4b}} \\
 \subfigure[]{
  \includegraphics[width=0.99\columnwidth]{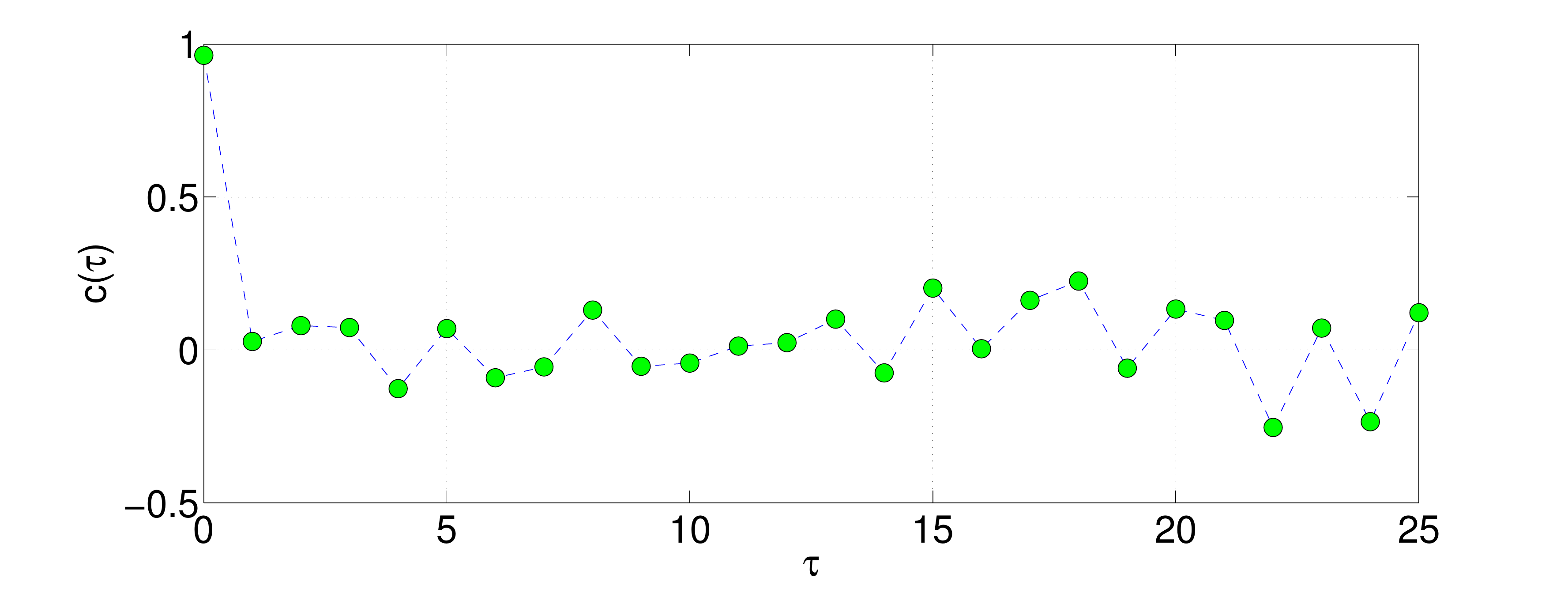}
  \label{fig4c}}
  \caption{\textit{Discretization of the process}. a) Example of
  a single continuous trajectory plotted versus
  time and the corresponding discretized position as green dots in
  the center of one single well oscillation. b) Corresponding discrete trajectory over 100 jumps, which
  conform to $10^4$ time units, as a function of the jump index
  $n$. c) Correlation function of the
  single trajectory from above as a function of time delay between
  two occurring jumps.}
  \label{fig4}
\end{figure}

\section{The quasi-random walk - numerical results}
Before obtaining insight from analytical considerations, we start by
simulating the dynamics of the system. This is achieved by fixing
the set of parameters to: $\Delta_a=-1.5 \kappa $, $\Delta_c=-1.5
\kappa$, $\eta_L=\kappa$, $\delta_T=\frac{\pi}{100}\kappa$, $\gamma=
\kappa$, $g=\kappa \times10^{-2}$, $k=2\pi$ and recoil frequency
$\omega_r=0.1 \kappa$. We treat $\eta_T$ as a free varying
parameter. In the following we set $\kappa=1$ for numerical
simulations and normalize the time in units of $\kappa^{-1}$. We
choose a regime described in the next analytical section as
'trapping via longitudinal pump', where we first tune the parameters
such that trapping of the particle is ensured in the absence of the
transverse driving. We then increase $\eta_T$, and notice that past
a given threshold, the particle starts jumping out of its trapping
site to the neighboring left/right sites in an apparently random
way. In Fig.~\ref{fig3} we exemplify such a trajectory obtained for
a particle initialized with $p_0=0$ around the origin at
$x_0=3.5\times10^{-3}$ and for $\eta_T=0.55 \kappa$.

\begin{figure}[b]
\includegraphics[width=0.85\columnwidth]{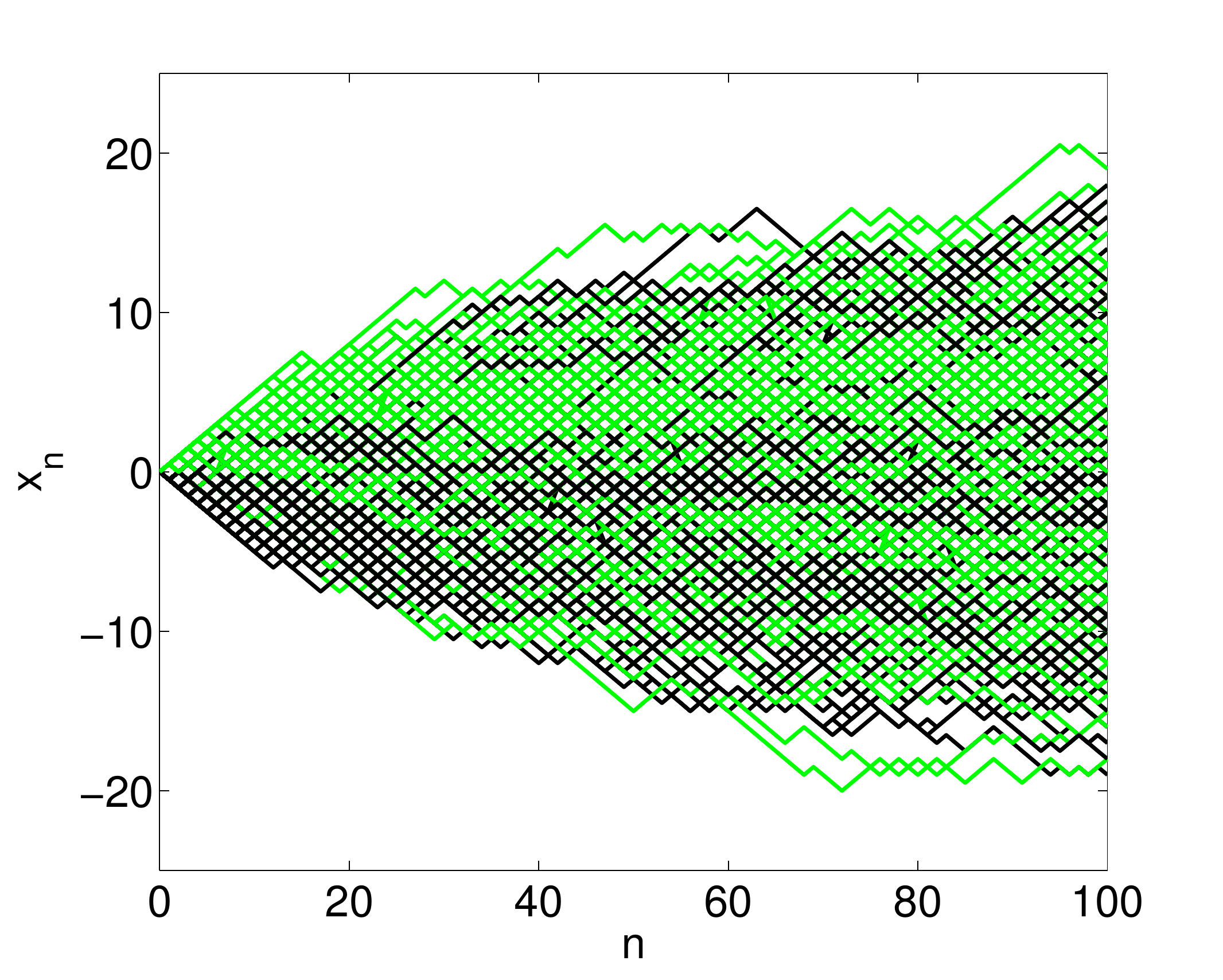}
\caption{\emph{Mixing of trajectories}. Simulation of $10^3$
discrete trajectories $x_n(x_0)$ as a function of the jump number
index $n$, starting with uniform distributed initial positions and
momenta out of the interval $[-0.1, 0.1]$. The color coding refers
to negative/positive initial positions (black/green): notice that
the evolution completely mixes the negative and positive regions.}
 \label{fig5}
\end{figure}

\begin{figure*}[t!]
  \subfigure[]{
   \includegraphics[width=0.60\columnwidth]{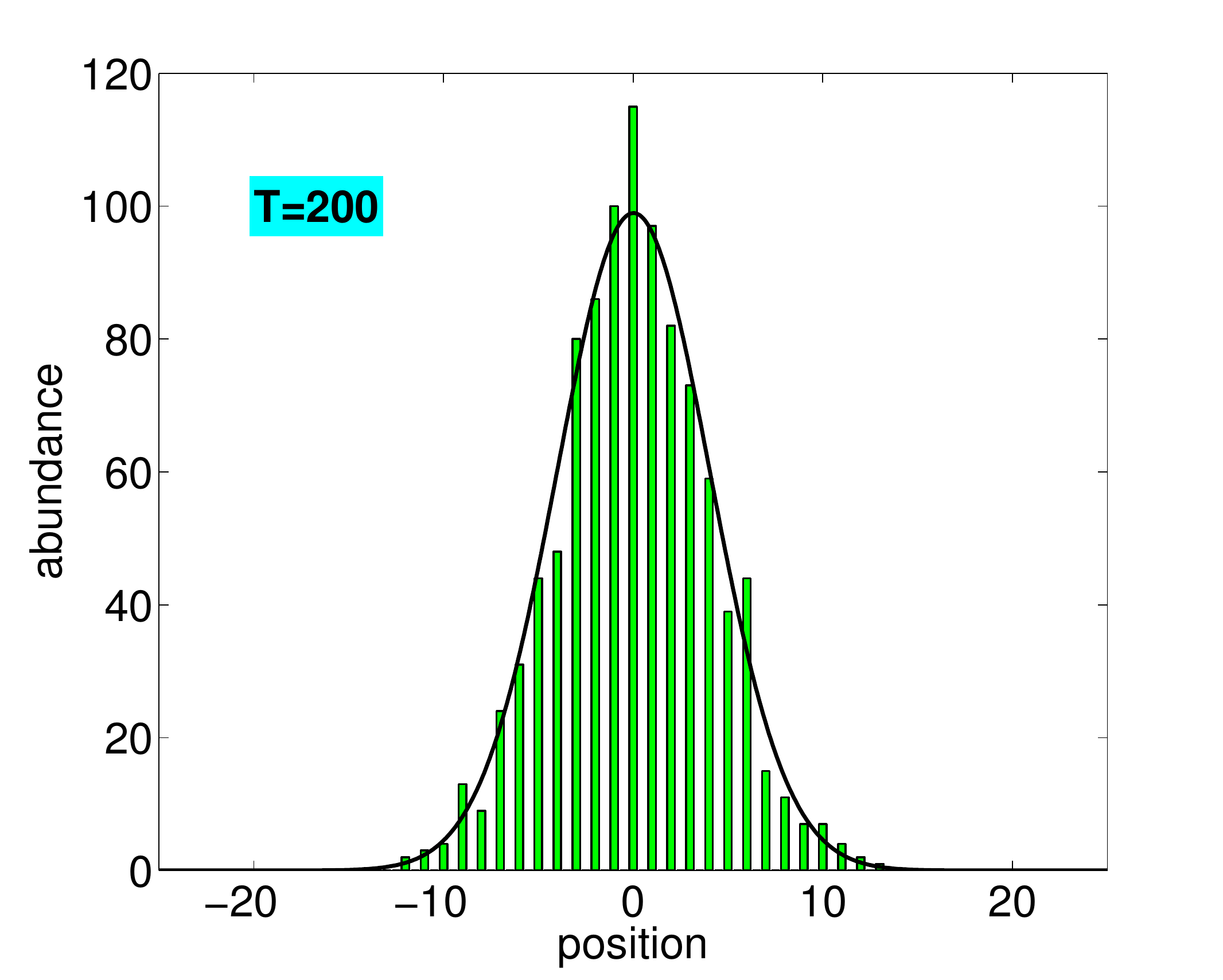}}
  \subfigure[]{
   \includegraphics[width=0.60\columnwidth]{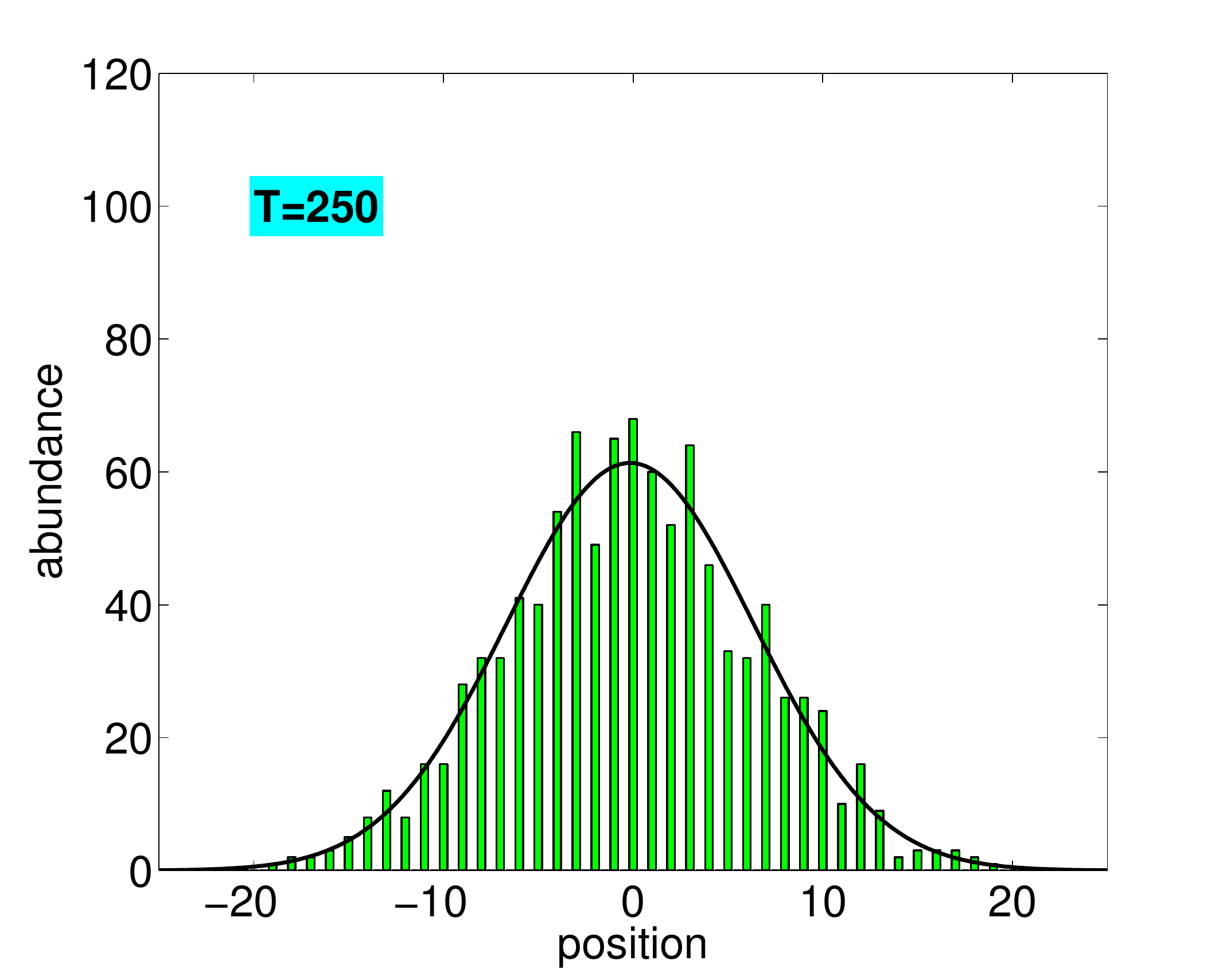}}
  \subfigure[]{
  \includegraphics[width=0.60\columnwidth]{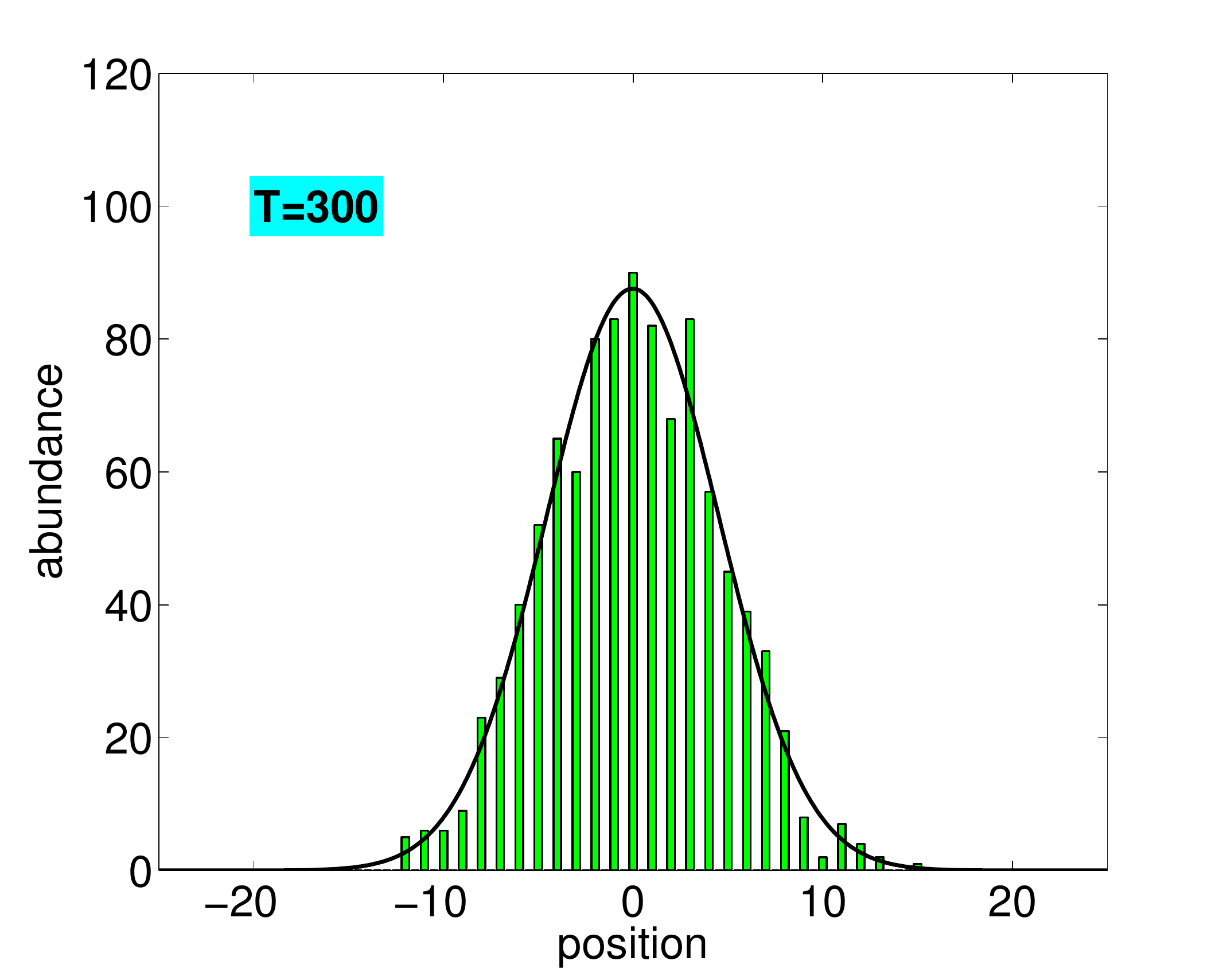}}
 \\
 \subfigure[]{
  \includegraphics[width=0.60\columnwidth]{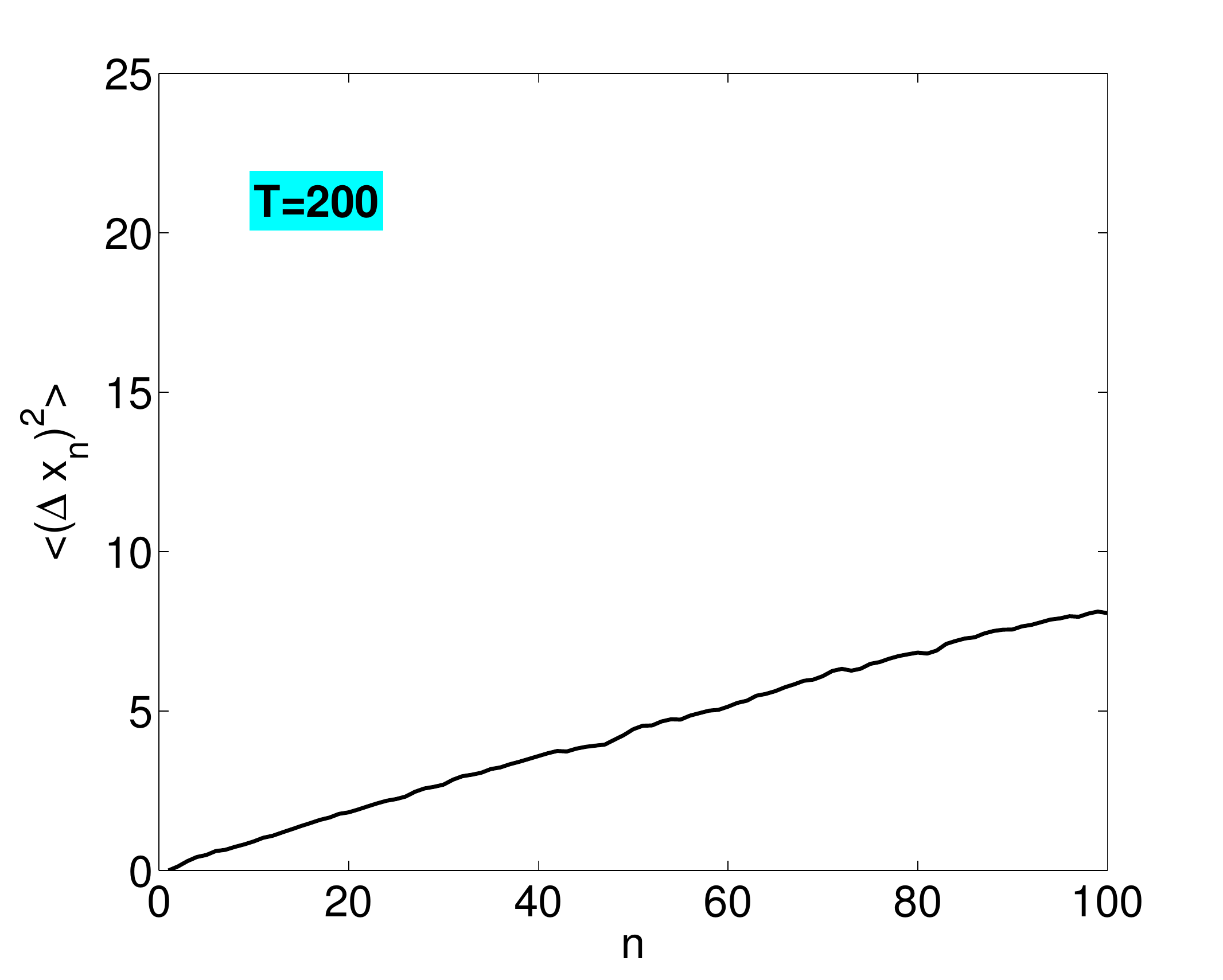}}
 \subfigure[]{
  \includegraphics[width=0.60\columnwidth]{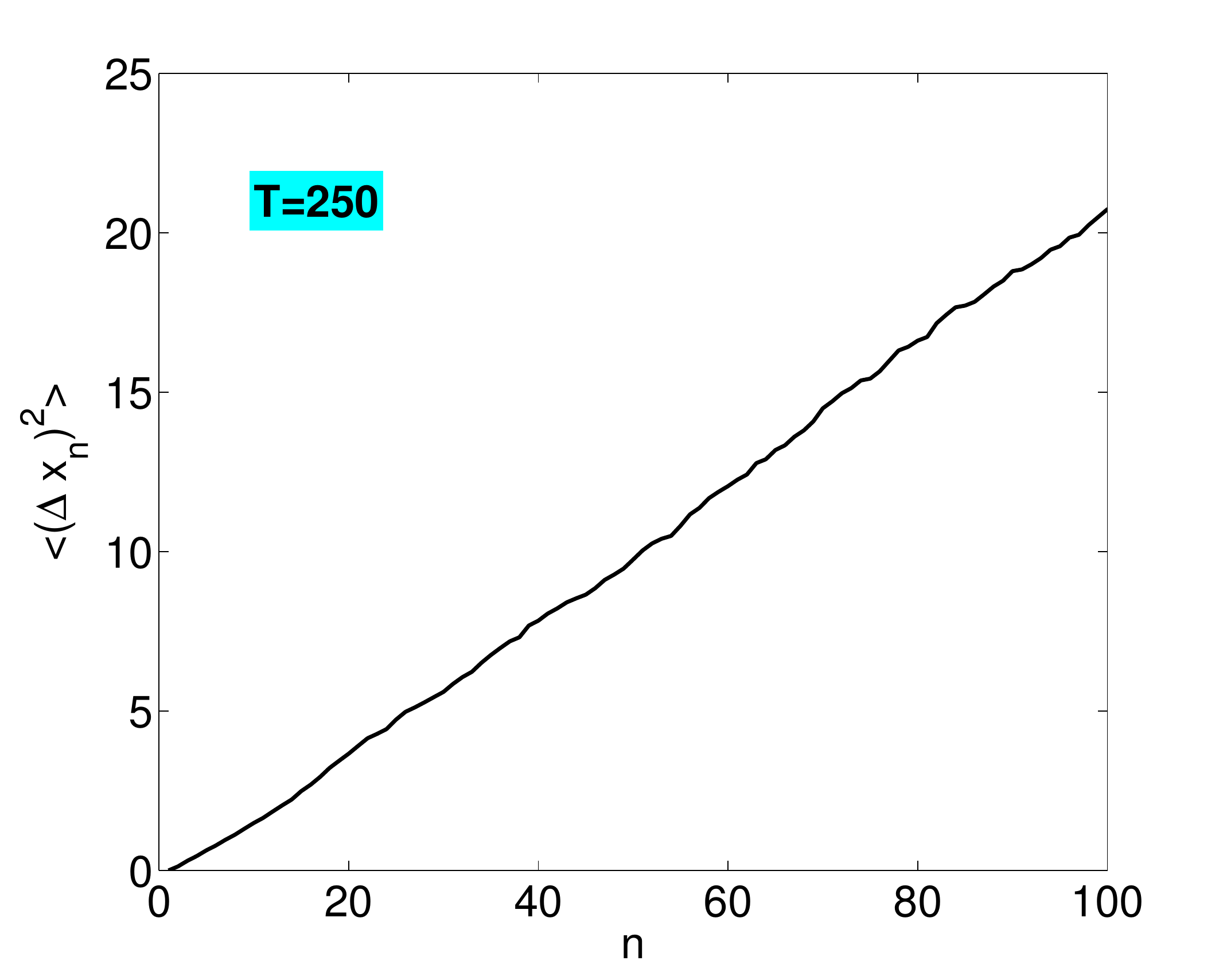}}
 \subfigure[]{
  \includegraphics[width=0.60\columnwidth]{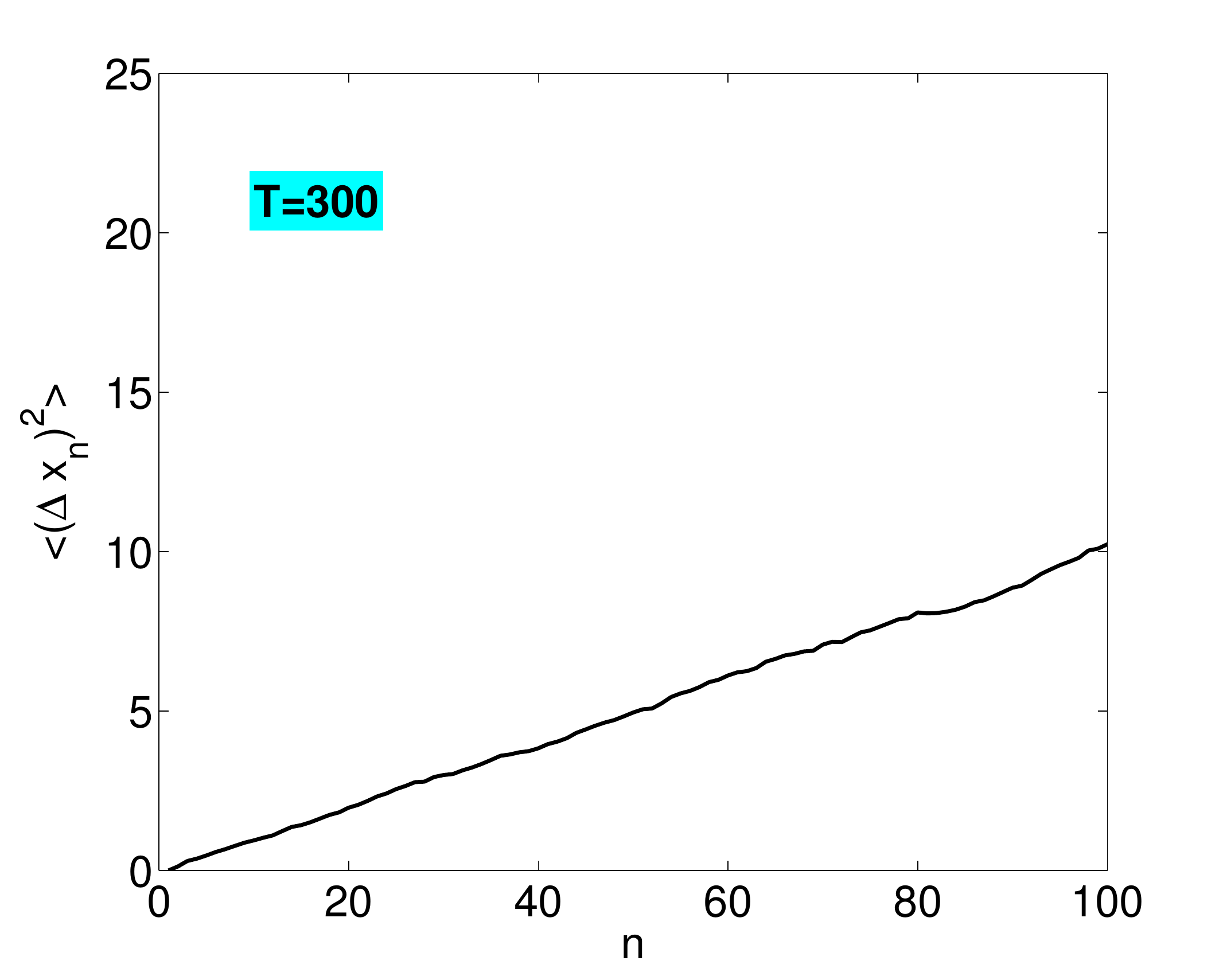}}
\caption{\emph{Final site occupancy and mean variances}.
a)-c) Histograms of populations of lattice sites at
arrival. The arising distributions fit well with the expected
binomial distribution (here the fit is with a Gaussian owing to the
large number of steps considered $N=100$) characteristic of a random
walk. The standard deviations of the distributions vary with the
time period $T$ between two jumps. This corresponds to different
diffusion constants, given by the slopes of the variance curves
below.
d)-f) Mean variance in position as a function of jump index
$n$. The linear increase with the amount of steps reproduces the
main property of a classical discrete one dimensional random walk.
For $T=250$ the slope is close to the expected value of $0.25$ for a
perfect random walk, while the other two cases show sub-diffusion.}
\label{fig6}
\end{figure*}

\subsection{Discrete process - single trajectories}
We then discretize the process by choosing time steps in units of
the time period between two jumps $T$, which is half the period of
the potential time oscillation. At $t_n=n T$ the particle is
released from the potential and jumps to an adjacent trapping site.
This is illustrated in Fig.~\ref{fig4} as the transition from the
continuous trajectories in the upper plot to the discrete plots of
site number in the middle plot. The discrete positions (the
locations of the sites) are defined as:
\begin{align}
    x_n\equiv \frac{1}{2T}\left[2\int_{t_{n-1}}^{t_n}dt x(t) \right]
\end{align}
where the square brackets stand for the rounding of the integral to
the nearest integer, corresponding to the location of
the site where trapping occurs.

With the parameters from above, we then analyze the dynamics as a
function of randomized initial conditions; we initialize the
particle with a momentum and position $(x_0,p_0)\in[-0.1,
0.1]\times[-0.1, 0.1]$ inside the potential well around zero and
follow the evolution over time normalized to $T$ for $10^3$ initial
values. First, we illustrate the mixing of trajectories, as shown in
Fig.~\ref{fig5}, by color coding the trajectories starting with
$x_0>0$ in green and those starting with $x_0<0$ in black. The
mixing is evident and can be taken as a first
indicator for randomness.\\
One can introduce the jump sequence $J_N=(j_1,j_2,...,j_N)$, where
the jump indicators are defined as $j_n=x_{n+1}-x_n$, and according
to their sign show either left or right jump behavior. One can
define the autocorrelation function for this process as
\begin{align}
C_{N}(\tau)&=\frac{1}{N-1}\sum_{n=1}^{N} (j_{n+\tau}-\langle
j_n\rangle ) (j_n-\langle j_n\rangle),
\end{align}
which characterizes the joint probability for the occurrence of
jumps separated by a time delay $\tau T$. The behavior of this
function for a single trajectory is shown in Fig.~\ref{fig4}(c) in
the lower plot. By definition the zero time delay correlation is
normalized to unity and it decreases to values around zero where
negative values show anti-correlated jumps while positive values
indicate correlated jumps.

\begin{figure*}[t!]
 \subfigure[]{
  \includegraphics[width=0.60\columnwidth]{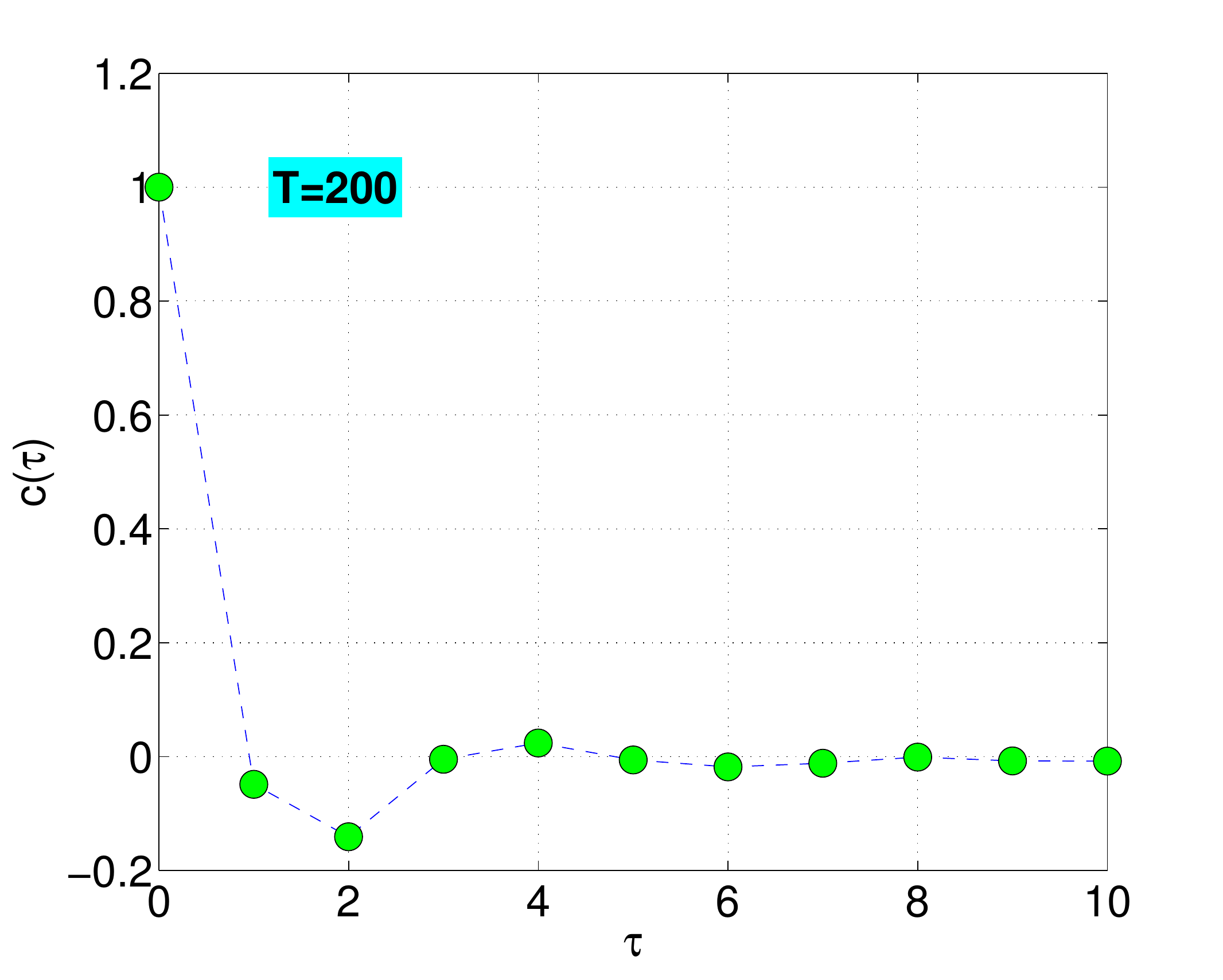}}
 \subfigure[]{
  \includegraphics[width=0.60\columnwidth]{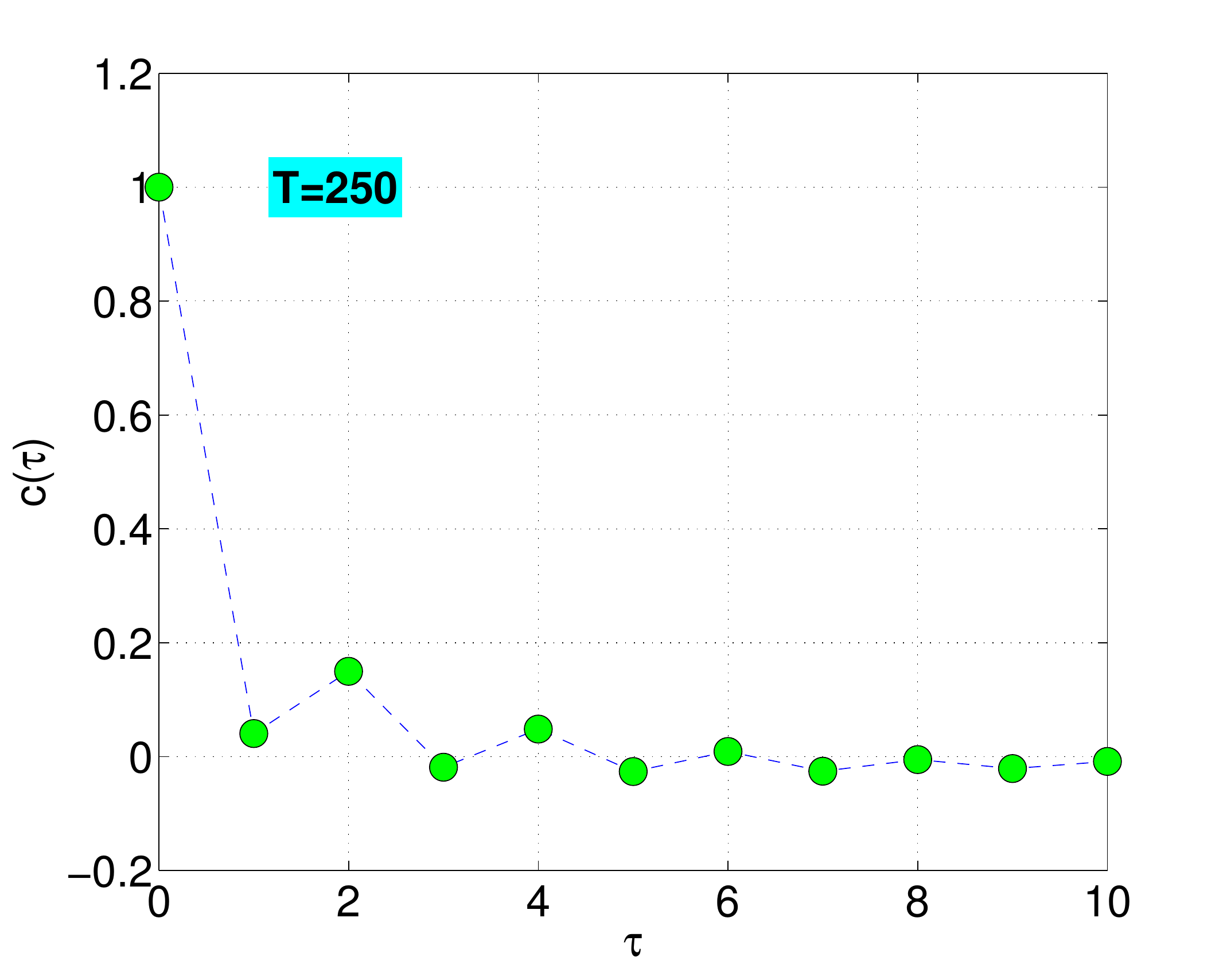}}
 \subfigure[]{
  \includegraphics[width=0.60\columnwidth]{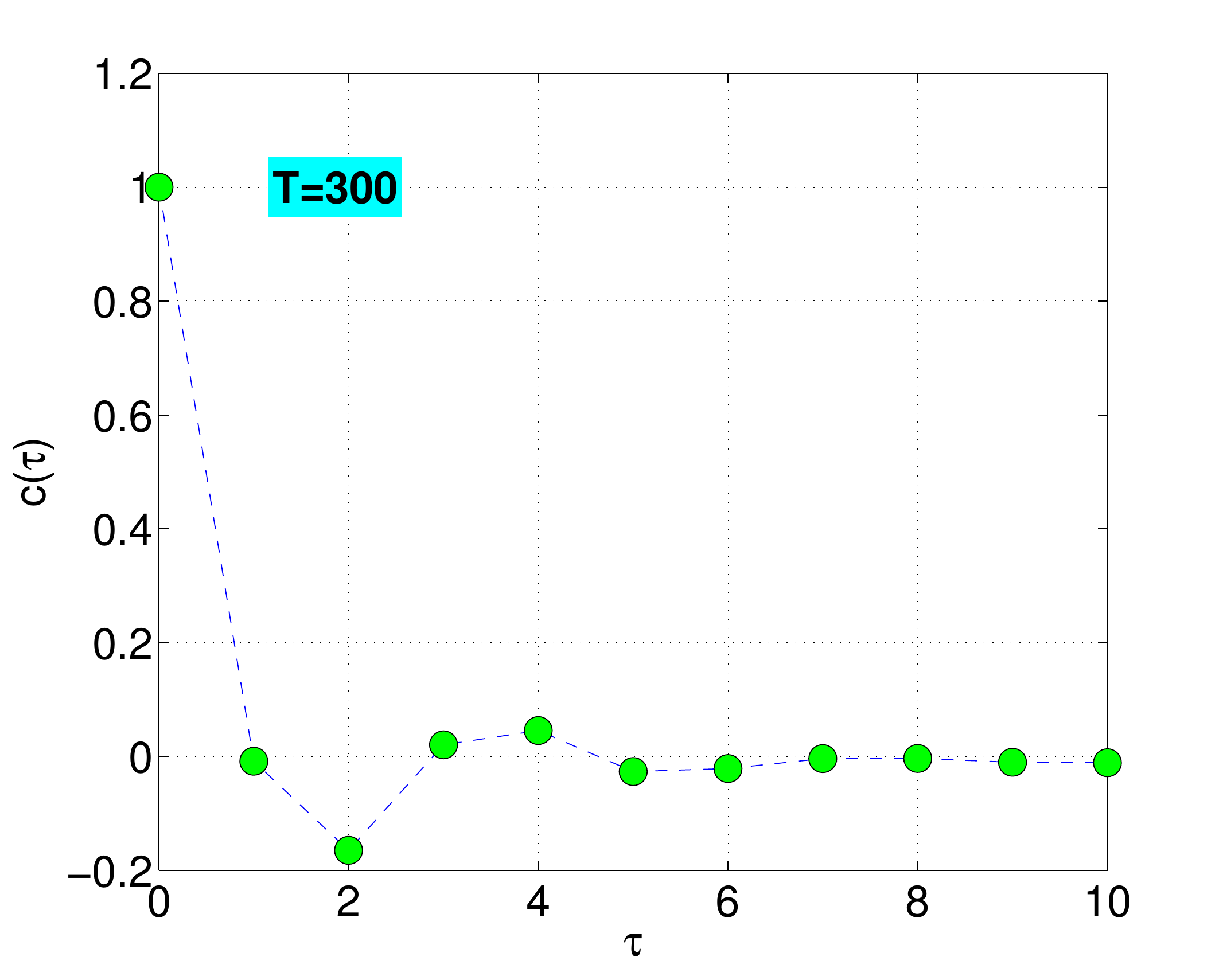}}
\caption{\emph{Averaged jump-correlations}. Numerical data showing
the variation of the correlation function with the delay time
between jumps for different values of $T$. As a basis of comparison,
the correlations of a pure random walk would correspond to a
function reaching unity for $\tau=0$ and zero elsewhere. On the
given numerical example, the particle is subjected to an effective
reservoir with a non-vanishing memory that allows for
anti-correlations of jumps close to each other. These strong
negative regions are missing in the middle case b) that corresponds to
the highest diffusion.} \label{fig7}
\end{figure*}

\subsection{Discrete process - many trajectories statistics}

We numerically simulate a large number of trajectories (with
starting point around the origin) for $10^2$ time steps and
different values of the jump period $T$.
The distribution of
the final site occupancy is illustrated in Fig.~\ref{fig6}a)-c).
The histograms of all three cases coincide with binomial distributions
expected from the classical random walk. The fitting is done with
a Gaussian distribution, where the standard deviation of the
histogram is considered as the distribution's width. The variance
in position $x$ of the unbiased binomial distribution is given by
$\left<(\Delta x_n)^2\right>=a^2n$, where $a$ is the constant
spatial separation of adjacent lattice sites. Since the trapping
positions in our model are separated by a distance of
$\frac{\lambda}{2}$ and we set $\lambda=1$, we expect a variance of
$\left<(\Delta x_n)^2\right>=\frac{1}{4}n$. The numerically
observed variances in Fig.~\ref{fig6}d)-f) show a
dependency of the slope of the linear increase, which is equivalent
to the diffusion constant, on the jump period $T$.

\noindent The main result of the numerical section is however the
behavior of the jump correlation function averaged over many
trajectories (see Fig.~\ref{fig7}). For a perfect random walk
process, the correlation function for jumps separated by $\tau>T$
would be vanishing. This corresponds to a reservoir having no
memory. In our case for $T=200$ and $300$ however, short time delays
(below 5 jumps) are anti-correlated while after around 5 jumps
correlations occur. These anti-correlations do not occur in the
second case, which shows the best coincidence with the expectation
for a perfect random walk. The anti-correlations seem to correspond
to sub-diffusion of the averaged particle motion as they favor
jumps back to origin at given time distances which might inhibit
the spread of the total motion.\\
To gain some physical understanding, one can inspect
Eq.~\eqref{eq:force1} where the right-hand side represents the
effective optical force. In some limit (revealed by the numerical
results) this force shows effective quasi-random kicks whose
correlations map onto the correlation function for the discrete
process. For perfectly uncorrelated kicks the effect would be a
random walk. However, in the realistic case some correlations
between jumps remain. One can consider the following argument: the
momentum kick at one site is the integral of the force over a time
$T$ during which the force varies non-trivially. In the continuous
limit the process is deterministic. However, in the limit of many
oscillations inside a single site, the phases of the momentum kicks
occurring at different sites are randomized. On the other hand the
numerical results show an alternating behavior of the system with
$T$. The fact that e.g. the diffusion constant decreases again from
$T=250$ to $T=300$ gives rise for the hypotheses that this system is
more close to a random walk for special ratios of the phases of the
single trapping site oscillations and the oscillation of the
potential in time.

\section{Analytical results}

The dynamics numerically derived in the previous section can be
explained at least in some particular limits by a simplified model
where the atomic and field degrees of freedom are eliminated and an
effective set of equations is derived for the particle motion only.
Let us first rewrite the effective force acting on the particle
[from Eq.~\eqref{eq:force1}] in the following form:
\begin{align}
   F(x,t)=& - 2 f'(x) \text{Im}\{ \alpha^{\ast} (g\beta)  \},
\end{align}
and proceed with finding analytical expressions in the adiabatic limit.

\subsection{Elimination of the atomic dipole}

Under the assumption of purely dispersive coupling brought by the weak, far
off-resonance driving $\Delta_a\gg \gamma ,\eta_T$ one can eliminate
the atomic variable and compute
\begin{align}
   \begin{split}
  g \beta  &  \simeq i \frac{g^2}{\Delta_a}\alpha f(x) + i \frac{g \eta_\tau
e^{i\delta_T t}}{\Delta_a}\\
    & = i U_0\alpha f(x)+ i \bar{\eta}_T e^{i\delta_T t},
    \label{beta_ss}
    \end{split}
\end{align}
where the per photon dispersive coupling is given by
$U_{0}=g^{2}/\Delta_a $ and the effective transverse pump is defined
as $\bar{\eta}_T=g \eta_T / \Delta_a$. Replacing the steady state
value of $\beta$ in the force expression we obtain:
\begin{align}
   F(x,t)=& -2f'(x)f(x) U_0 |\alpha|^2 - 2f'(x) \bar{\eta}_T \text{Re}\{ \alpha^* e^{i\delta_T t} \} .
   \label{force2}
\end{align}
\noindent Notice that the first term is the well-known force
arising from the longitudinal pump into the cavity. Combined with
time-delay effects coming from the finite ring-down time of the
cavity field, such a force can lead to cavity
cooling, heating, bistability or self-oscillations~\cite{Hechenblaikner1998Cooling,Ritsch2013Cold}. The second
term is of more importance in our treatment as it shows the
interference between the two pumps and it contains the
time-modulation needed for the potential sign change.

\subsection{Elimination of the field variable}

Since the time-delay effects do not play a role in the occurrence of jumps, we proceed by considering the limit of small $U_0$
where we eliminate the cavity field. Replacing $\beta$ from
Eq.~\eqref{beta_ss}, leads to

\begin{align}
    \dot{\alpha} & = \left[-\kappa+i (\Delta_c-U_0 f^2(x)\right] \alpha + \eta_L - i\bar{\eta}_T f(x) e^{i\delta_T t}.
\end{align}

\noindent Under the assumption that $\eta_L,
\bar{\eta}_T\ll \text{max}(\kappa,\Delta_c)$ and defining a position dependent cavity detuning $\Delta_c-U_0 f^2(x) \equiv \Delta(x)$, one
obtains:
\begin{align}
   \alpha = & \frac{\eta_L-i \bar{\eta}_T f(x) e^{i\delta_T t}}{\kappa-i \Delta(x)}.
\end{align}

\noindent We now can compute the cavity photon number
\begin{align}
   |\alpha|^2 = & \frac{\eta_L^2+ {\bar{\eta}_T}^2 f(x)^2+ 2 \eta_L \bar{\eta}_T f(x) \sin(\delta_T
   t)}{\kappa^2+\Delta(x)^2},
\end{align}
as well as
\begin{align}
   \text{Re}\{ \alpha^* e^{i\delta_T t}  \} = & \frac{\eta_L\left[\kappa \cos{\delta_T t}+\Delta(x) \sin(\delta_T t)\right]+ \Delta(x) f(x) \bar{\eta_T}}{\kappa^2+\Delta^2(x)},
\end{align}
and replace these expressions in Eq.~\eqref{force2}.

\subsection{Optical forces}

We can now group the different terms contributing to the total
optical force acting on the particle as follows: i) arising from the
longitudinal field, ii) from the transverse field and iii) a time
dependent interference term. Explicitly writing the three terms as
$F(x,t)=F_{L}(x)+F_{T} (x)+F_{LT}(x,t)$, we compute
\begin{align}
   F_L=& -2f'(x)f(x) \frac{\eta_L^2}{\kappa^2+\Delta^2(x)} U_0,
\end{align}
which describes the standard $\cos^2(x)$ optical potential induced
by the longitudinal pump. The next term is
\begin{align}
   F_T=&-2f'(x)f(x) \frac{ \bar{\eta}_T^2 }{\kappa^2+\Delta^2(x)} \Delta_c,
\end{align}
and it shows the effect of the time-independent interference between transverse pump photons and the particle scattered photons filling the cavity mode.
The most interesting term is
\begin{align}
   F_{LT}=& 2f'(x) \frac{ \bar{\eta}_T \eta_L}{\kappa^2+\Delta^2(x)} [\kappa \cos(\delta_T t)  \nonumber \\
          & +(\Delta_c+U_0 f^2(x)) \sin(\delta_T t)],
\end{align}
showing modulation in time at $\delta_T$. Notice that the time
independent limit can be reached by setting $\delta_T=0$, and this
force reduces to $ 2f'(x) \bar{\eta}_T \eta_L
\kappa/(\kappa^2+\Delta^2(x))$.

\subsection{Trapping by interference}
Let us consider the limit of small $g$ and tune the driving field
amplitudes such that $\eta_T\gg g \eta_L$ while at the same time
$\eta_T\ll\eta_L \Delta_a /(g \Delta_c)$. To satisfy both conditions
simultaneously one has to require $U_0 \Delta_c\ll 1$. Under these
conditions, the time interference force is dominant and gives rise
to an effective time-modulated trapping potential. We neglect the
spatial modulation $U_0 f^2(x)$ as well with respect to the larger
$\Delta_c$ and obtain the total force on the particle simplified to:
\begin{align}
   F \simeq & -2\sin (x) \frac{ \bar{\eta}_T \eta_L \kappa}{\kappa^2+\Delta_c^2} \cos(\delta_T
   t-\phi_\Delta),
   \label{forceinterference}
\end{align}
where the phase is defined as: $\phi_{\Delta}=\arctan{\frac{\Delta_c}{\kappa}}$.
The equations of motion for the particle are those of a frequency
modulated pendulum and similar to the ponderomotive force with an
important difference in that the amplitude of the driving changes
sign periodically. In the limit of good localization (where we can
expand $\sin(x)\simeq x$) we identify this as a restoring force for
an harmonic oscillator with a time dependent normal frequency. We can
readily compute the maximum trapping frequency $\omega_{tr}$ as
\begin{align}
   \omega_{tr,LT}^2 = & 4 \omega_r \kappa \frac{ \bar{\eta}_T
   \eta_L}{\kappa^2+\Delta_c^2}.
\end{align}
The quasi-random walk behavior arises from the periodic force sign
change which, after a number of oscillations at a given site
(roughly proportional to $T/T_{tr,LT}$ where
$T_{tr,LT}=2\pi/\omega_{tr,LT}$) forces the particle to settle
itself inside an adjacent well.

\subsection{Trapping by longitudinal pumping}
A different regime is reached when the $F_L$ and $F_{LT}$
contributions are of the same order of magnitude. We achieve this by
first turning on the longitudinal pump and see that a trapping
time-independent potential is established. Along the equilibrium
points, the longitudinal trapping frequency is around:
\begin{align}
   \omega_{tr,L}^2 = & 8 \omega_r \frac{U_0 \eta^2_L}{\kappa^2+\Delta_c^2}.
\end{align}
Tuning up the transverse pump, a time modulation of the trap
frequency is achieved and a threshold emerges for $\eta_T$ (for a
rough estimate one can equate the maxima of $F_L$ and $F_{LT}$)
after which the particle starts jumping to adjacent sites. Given the
two different spatial modulations of the forces, $\sin(2x)$ vs.
$\sin(x)$, a double well structured potential arises with
frequencies:
\begin{align}
  \bar{ \omega}_{tr,L}^2 = & 8 \omega_r \frac{\eta_L(U_0 \eta_L\pm \eta_T)}{\kappa^2+\Delta_c^2}.
\end{align}

\subsection{Brownian motion for non-delta correlated forces}
We have already observed that a low diffusion constant is connected
to oscillations in the correlation function $C(\tau)$, namely
negative parts showing anti-correlations of jumps. While in the
previous sections we discussed the discrete random walk process, we
move now to the continuum and test that for a process undergoing
brownian motion, non-delta correlated forces can indeed give rise to
sub-diffusive behavior (i.e. the convergence of the variance in
space as function of time to a linear function with a slope below
the solution for a delta-correlated force). The ansatz that we take
for the expressions of  the force-correlations is suggested by the
shape of the jump-correlation functions of the discrete process. We
start by assuming a particle underlying the stochastic equation of
motion
\begin{align}
\frac{dv}{dt}=-\lambda v+\xi(t).
\end{align}
Following the derivation in~\cite{Lax2006Random} the variance in
space is
\begin{align}
\left<(\Delta
x)^2\right>=\int_0^tdt'\int_0^tdt''\phi(t')\phi(t'')\left<
\xi(t')\xi(t'')\right> \label{eq66}
\end{align}
with $\phi(t')=\frac{1}{\lambda}(1-e^{\lambda(t'- t)})$ and the
analytical solution for
$\left<\xi(t')\xi(t'')\right>=2d\delta(t'-t'')$ is
\begin{align}
\left<(\Delta x)^2\right>=\frac{2d}{\lambda^2}(t-\frac{2}{\lambda}
(1-e^{-\lambda t})+\frac{1}{2\lambda}(1-e^{-2\lambda
t}))\label{eq67}.
\end{align}
For large times (or large $\lambda$) the expression above converges
to the expected linear dependence $\left<(\Delta
x)^2\right>=\frac{2d}{\lambda^2}t$. In our units, this corresponds
to the typical linear diffusion with a slope of $1/4$. We now
compare this solution to the numerically calculated variance for a
force-correlation function of Gaussian and exponential shape times a
cosine modulation for different oscillation frequencies. The
expressions that we test are
\begin{align}
\left<\xi(t')\xi(t'')\right>=2d \left\{
    \begin{aligned}
        &e^{-(t'-t'')^2/2\sigma^2}\\
        &e^{-|t'-t''|/\sigma}
    \end{aligned} \right\}
        \cos(\Omega(t'-t'')).
\label{eq610}
\end{align}
We keep the width for all test functions fixed on a value of
$\sigma=0.5$ and vary the oscillation frequency $\Omega$ from $0.5$
to $2.0$ in steps of size $0.5$.

\begin{figure}[t]
\includegraphics[width=0.95\columnwidth]{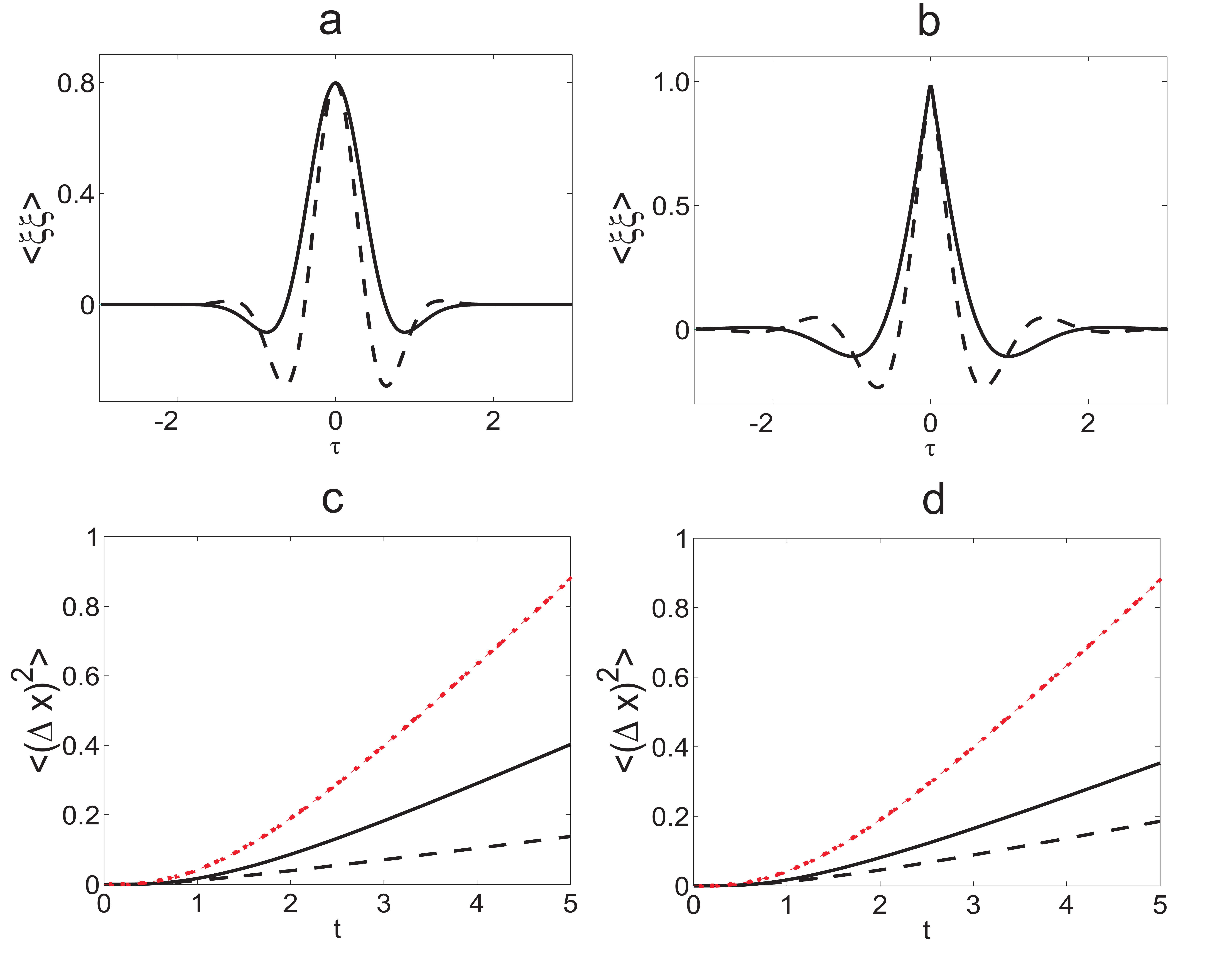}
 \caption{\emph{Force correlation functions and numerical solutions
of the variance.} a)-b) Correlation as a function of time delay is
illustrated for width $\sigma=0.5$ and frequencies $\Omega=0.5$
(black, full line) and $2.0$ (black, dashed) for the Gaussian (left)
and the exponential (right) case. The values of the local minima
become lower as $\Omega$ increases and expand the segments of
anti-correlation. c)-d) Variances (same color pattern as above)
converging to linear functions with slopes below the analytical
solution (red dashed line).} \label{fig8}
\end{figure}

\noindent As seen in Fig.~\ref{fig8}, the anti-correlations for this
continuous process drop the diffusion of the particle's motion
drastically and thus lead to sub-diffusion as expected. One should
consider the fact that in this scenario the correlation functions do
not converge to delta functions in the limit of vanishing width and
frequency since they are not normalized.

\section{Discussions}

\subsection{Frequency comb driving}
In the limit where the pump interference provides the sole trapping
mechanism, the force acting on the particle as in
Eq.~\eqref{forceinterference} reminds of a parametrically driven
pendulum. The dynamics consists of pendulum motion inside a given
trap (which reduces to harmonic motion in the limit of good
localization) with a time modulated frequency. However, one can
engineer a multiple frequency transversal pumping scheme where the
time modulation becomes a periodic kick instead of a sinusoidal
function. To this end we consider a frequency comb driving (with
$2N_f+1$ frequencies) with the minimum frequency separation $\delta$
and centered at $\omega_L$, such that the pump frequencies are
$\omega_{T,j}=\omega_L-j \delta$. If we neglect spontaneous atomic
decay and the coupling between atom and cavity field,
Eq.~\eqref{eqalphabeta} with the total pump term inserted gives
\begin{align}
  \dot{\beta}(t) \simeq i\Delta_a\beta +
                    \eta_T \sum_{n=-N_f}^{N_f} e^{i n \delta t}.
                    \label{betacomb}
\end{align}
In the limit $N_f\rightarrow \infty$ the driving on the right-hand
side becomes a Dirac comb function. Inserting the steady state
solution of Eq.~\eqref{betacomb} together with $\alpha \simeq\eta_L$
into Eq.~\eqref{eq:force1} we get an equation of motion that maps
onto the kicked rotor dynamics:
\begin{align}
 \dot{p}\simeq - \frac{\eta_L \bar{\eta}_T}{T_{\delta}}\sin(x)\sum_{n \in \mathbb{Z}}
    \delta(t-n\text{T})
\end{align}
where $T_{\delta}=2\pi/\delta$. Together with Eq.~\eqref{eq:force1}
the system can be exactly described via a transformation to the
discrete and can be shown to exhibit chaotic motion past a threshold
characterized by the tuning up of the force amplitude factor.

\subsection{Hybrid optomechanics with doped nano-spheres}

We can extend our treatment to consider a hybrid optomechanical
system where we replace the two level system with a doped nano-sphere containing a collection of
$N$ such systems. We assume the nano-sphere transparent to light except for the doped part where the cavity mode excites a transition close to resonance. Let us consider the nano-particle of mass $M$ with radius
much smaller than specific length in which the cavity mode changes
considerably (the cavity mode wavelength). The light-matter interaction takes place
via the Tavis-Cummings Hamiltonian, that changes from the single
atom picture in that $\hat{\sigma}^\pm$ is replaced by
$\sum_{j}\hat{\sigma}_j^\pm\equiv S^\pm$. In the bosonic limit,
where the saturation is very low, we can assume that
$\left[S^-,S^+\right]=-N$ and proceed to write equations for
averages ($\beta_N=\langle\hat{S}^-\rangle$):
\begin{align}
\dot{\alpha } &=(-\kappa +i\Delta_c)\alpha
                -g f(x)\beta_N +\eta _L, \\
\dot{\beta_N }  &=(-\gamma +i\Delta_a)\beta_N
                +g N f(x)\alpha  + N \eta _T e^{i\delta_T t}.
\end{align}
The immediate gain in this approach from the single atom approach is
the relaxation of the requirement of $|\beta|^2 \ll 1$ that turns into
$|\beta_N|^2 \ll N$. However one has to pay the price of a reduced recoil frequency
owing to an increased mass at least $N$ times larger. The upshot is that a cavity QED regime, where we would expect a quasi-random walk with a
macroscopic particle, can be unraveled.

\subsection{Implementation considerations}
To experimentally observe the proposed quasi random walk, we advance
a possible two step experimental procedure: i) turn on longitudinal
pumping detuned by $-\kappa$ from the resonance such that cavity
cooling takes place in the absence of transverse pumping and ii)
turn on $\eta_T$ past the threshold such that jumps are initiated.
The typical cooling procedure can ideally cool the particle towards
a thermal wavepacket with minimum energy $\hbar \kappa$ divided
between the position and momentum quadratures (according to the
thermal equipartition principle). In terms of normalized momentum
and position initial variations this corresponds to $\delta
p_0=1/\sqrt{2\omega_{r}}$ and $\delta x_0=1/(\sqrt{2U_0} \eta_L)$.
For localization within a site notice that we require $\delta x_0\ll
1$ which results in $U_0 \eta^2_L\gg 1$. This can be still fulfilled
in the polarizable particle regime by tuning $\eta$ such that
$g^2/\Delta_a \ll 1$ while $g \eta_L \gg 1$.

\section{Conclusions and outlook}

We considered the dynamics of particles (either as single two-level
systems or sub-micron spheres doped with multiple emitters) inside
time-dependent potentials resulting from interference in a
two-color, two directional pump scheme. Past a given threshold,
chaotic-like behavior can be observed, with correlations very close
to those of a typical random walk. Depending on the beat frequency,
the particles motion exhibits sub-diffusion that can be related
to anti-correlations of the acting force. Analytically the system in
the steady state can be reduced to a pendulum with a time dependent
frequency modulation.

\noindent While the treatment here is in
the classical regime, an immediate generalization into the quantum
realm can be made by either: i) treating motion classically and
considering the effect of the quantum nature of the two-level system
onto the dynamics or ii) treating motion quantum mechanically and
analyzing the dynamics of an initial wave packet, with direct
connection to matter-wave interferometry applications.
Another future direction aims to extend the 1D treatment
to 3D dynamics where the beating of the two pumps give rise to an
effective ponderomotive force. Investigations will be carried out on
the possibility to exploit such a force for all optical trapping of
polarizable particles (or realistic multilevel atoms) inside 3D
optical cavities, similar to the mechanism employed in ion trapping
inside linear Paul traps.

\section {Acknowledgments} We thank Tobias Griesser for inspiring discussions. We acknowledge support from the Austrian Science Fund (FWF)
via project P24968-N27 (C.~G.) and via the SFB Foqus project F4013
and I1697-N27.

\end{document}